# Photonic Cluster State Generation from a Quantum Dot Emitting in the Telecom C-band

Giora Peniakov[1], Reza Hekmati[1], Johannes Michl[1], Mohamed Helal[1], Moritz Meinecke[1], Jochen Kaupp[1], Yorick Reum[1], Martin Kamp[1], Monika Emmerling[1], Andreas Theo Pfenning[1], Sven Höfling[1], and Tobias Huber-Loyola[1,2]

[1]Julius-Maximilians-Universität Würzburg, Physikalisches Institut and Würzburg-Dresden Cluster of Excellence ctd.qmat, Lehrstuhl für Technische Physik, Am Hubland, 97074 Würzburg, Deutschland
[2]Karlsruhe Institute of Technology, Institute of Photonics and Quantum Electronics, 76131 Karlsruhe, Germany

*giora.peniakov@uni-wuerzburg.de

Abstract

**Photonic cluster states are a key resource for photonic quantum information processing. So far, deterministic generation of these states has been limited to the near-infrared wavelength range. To achieve quantum advantage in communication while maintaining compatibility with silicon photonics, operation in the telecom wavelength range is required. In this work, we demonstrate deterministic cluster state generation directly in the telecom C-band. This is achieved through repetitive excitation of a hole spin confined in an indium-arsenide quantum dot subjected to an external magnetic field. We characterize the quantum process that generates the cluster state by measuring its process map, obtaining a fidelity of $\mathcal{F} = 0.71 \pm 0.01$ to the ideal case. As part of this characterization, we observe spin–photon polarization entanglement with a negativity of $\mathcal{N} = 0.27 \pm 0.02$. The emitted photons exhibit indistinguishability of at least 83%, demonstrating the potential for future fusion gates necessary for photonic cluster state generation beyond linear connectivity.**

Photonics is widely regarded as a promising platform for quantum computing and quantum communication [1–3]. For both applications, the multipartite entangled photonic cluster states are a key resource. In photonic quantum computing, such states form the basis of the measurement-based approach. When combined with feed-forward operations, they enable deterministic computation and provide an alternative to the circuit model [4]. Their compatibility with integrated on-chip silicon photonic platforms supports the scalability of photonic quantum computing [2,5,6]. In quantum communication, cluster states offer an alternative for building quantum networks without the need for quantum memories by offering an all-photonic approach [7–11]. Existing optical telecommunication networks are based on silica fibers that exhibit minimal transmission loss at wavelengths around 1.55 μm. Consequently, direct generation of photonic cluster states in this wavelength range would enable straightforward integration with existing fiber infrastructure for quantum communication applications without the need for frequency conversion [12–15].

One approach to generate cluster states is by fusing single photons or entangled photon pairs into larger cluster states using linear optics [16–20]. Photon generation for this purpose can be achieved at the telecom wavelength range either via spontaneous parametric down-conversion in nonlinear crystals [21], or deterministically from semiconductor quantum dots [22–24]. However, the required fusion operations are inherently probabilistic, limiting the scalability of this approach toward large photonic cluster states [2,3]. An alternative approach is to generate cluster states deterministically from single photon sources, using variants of the Lindner and Rudolph scheme [25]. In these platforms, a stationary two-level system sequentially emits photons directly in a cluster-state, without requiring additional entangling operations. Recent demonstrations achieved record-long cluster states of up to 14 photons using single atoms [26,27], while quantum dots enabled generation rates orders of magnitude faster, reaching the GHz regime [28–34]. However, so far, these demonstrations were performed in the near-infrared wavelength range around 780–960 nm. Converting such photons to telecom wavelengths requires additional frequency-conversion techniques with limited efficiency, introducing similar scalability constraints [12–15].

Here, we overcome these limitations by generating photonic cluster states directly in the telecom C-band using a quantum dot emitter operating at 1549 nm. Following the Lindner–Rudolph protocol, we employ a confined hole spin that is periodically excited to sequentially emit entangled photons. In combination with an externally applied magnetic field, this enables direct generation of photonic cluster states. The generated photons exhibit high indistinguishability (83%), supporting future fusion operations toward higher-dimensional cluster state architectures [3,35,36]. During the submission process of this manuscript, we became aware of a related work on cluster state generation at telecom using a similar scheme [37].

We perform our measurements on an $InAs/In_{.53}Al_{.23}Ga_{.24}As$ quantum dot embedded in a deterministically placed circular-Bragg grating microcavity [38,39] (**Fig. 1**a). A photoluminescence (PL) spectrum of the quantum dot under above-band-gap excitation is shown in **Fig. 1**b. The $X^+$ spectral line appears at 1549.1 nm, at the center of the telecom C-band. It corresponds to the optical transition between the positive trion, consisting of two holes and an electron, and a single hole.
Despite the spectral detuning between the $X^+$ and the cavity mode (centered at 1560 nm with a FWHM of 10 nm), the cavity induces a Purcell enhancement reducing the radiative lifetime of the $X^+$ to $\tau_L = 238 \pm 5$ ps [38]. The cluster state generation protocol requires an excitation scheme that enables qubit encoding in photon polarization. For that, a quasi-resonance located 10.2 meV above the $X^+$ emission is used, corresponding to an excited state of the trion, $X^{+*}$. In this state, one of the comprising charge-carriers is excited to a higher confined energy level, which is most likely a hole excited to the p-shell. This quasi-resonance preserves the circular selection rules of the hole–$X^+$ transition, so that excitation with circularly polarized light leads to strongly co-polarized emission. The co- to cross-polarized emission ratio exceeds 10:1, indicating decent "polarization memory" (see SI, Section 7).

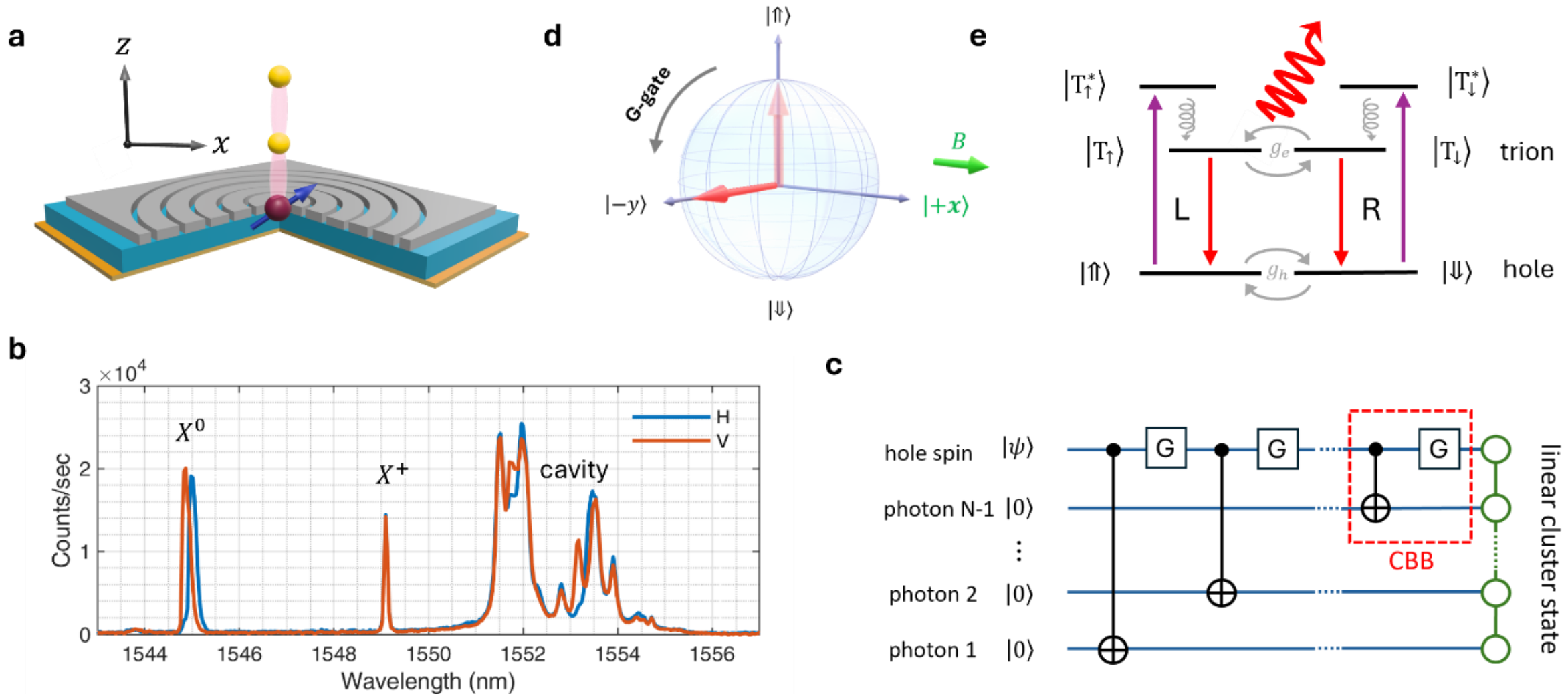

**Fig. 1. a** Circular-Bragg grating microcavity. **b** Polarization resolved photoluminescence under above-band excitation. $X^0$ is the neutral exciton and $X^+$ is the positive trion recombination transition. The broad emission feature around 1552 nm comes from other transitions feeding into the cavity. H(V) stands for horizontal(vertical) linear polarization. **c** Cluster-state generation circuit consisting of CNOT and $G$ gates. Given an input state $|\psi\rangle = |-y\rangle$, a linear cluster state is generated. A red dashed frame marks a single circuit building block (CBB) that we characterize using a quantum process tomography. **d** Bloch sphere representation of the ground state hole spin illustrating the quarter-precession $G$ gate. **e** Selection rules of the hole–$X^+$ optical transition enabling the CNOT gate. Curley grey arrows indicate phonon relaxation, which preserves spin orientation.

The underlying protocol to deterministically generate a cluster state is schematically shown using the quantum circuit presented in **Fig. 1**c. The top wire of this circuit is associated with a hole spin acting as a photon entangler. Its spin projections along the optical axis $z$, $|\Uparrow\rangle$ and $|\Downarrow\rangle$, encode the qubit states $|0\rangle$ and $|1\rangle$, respectively. At every application of a controlled-NOT (CNOT) gate, a photon is generated with polarization that encodes an additional qubit in the circuit. In between the CNOT gates applied are single qubit gates $G = \exp(-i\sigma_x\pi/4)$ (see **Fig. 1**d), resembling the operation of the Hadamard gates in the original proposal [25]. When the input spin qubit is initialized in a state $|\psi\rangle$ that is an equal superposition of $|\Uparrow\rangle$ and $|\Downarrow\rangle$, the circuit outputs a multi-qubit state that is locally equivalent to a linear cluster state.

To implement the $G$ gate, we apply an external magnetic field along the $x$-axis (see Fig. 1a), which induces Larmor precession of the spin around this axis. The $G$ gate is executed by allowing the spin to precess for one quarter of the precession period, $T_h$, which is determined by the hole $g$-factor, $g_h = 0.37(2)$, and magnetic field strength. In our experiment, we set the magnetic field to 50 mT, corresponding to $T_h = 4.2$ ns.
The CNOT gate is implemented using the optical transition between the hole and the $X^+$ (**Fig. 1**e). The control qubit of the gate is encoded in the hole spin and is promoted to the $X^+$ spin upon excitation. Within its lifetime, the system decays back to the hole while emitting a photon whose polarization encodes the target qubit. Specifically, we utilize the circular selection rules of the $X^+$-hole transition, $|T_\uparrow\rangle \leftrightarrow |\Uparrow\rangle|L\rangle$; $|T_\downarrow\rangle \leftrightarrow |\Downarrow\rangle|R\rangle$. Here, $|T_\uparrow\rangle$ and $|T_\downarrow\rangle$ denote the $X^+$ spin projections determined by the unpaired electron in the trion, while L and R denote left- and right-handed circularly polarized light. Since in the circuit the CNOT and $G$ are always applied consecutively, a single sequence of the two constructs the circuit building block (CBB) (see **Fig. 1**c). Therefore, to experimentally characterize the circuit operation on $N$ qubits, it is enough to characterize a CBB only once and then concatenate the result $N-1$ times [28]. For this, we utilize quantum process tomography; we prepare the input spin in four linearly independent states $|\psi\rangle \in \{|+x\rangle, |\pm y\rangle, |+z\rangle\}$, and measure the corresponding output spin–photon states $\rho(|\psi\rangle)$. In this notation, $|+z\rangle = |\Uparrow\rangle$, $|-z\rangle = |\Downarrow\rangle$, $\sqrt{2}|+x\rangle = |\Uparrow\rangle + |\Downarrow\rangle$, $-\sqrt{2}i|-x\rangle = |\Uparrow\rangle - |\Downarrow\rangle$, and $\sqrt{2}|\pm y\rangle = |\Uparrow\rangle \pm i|\Downarrow\rangle$. From the four input–output measurement pairs, we calculate a $4\times4\times4$ tensor $\Phi$, completely mapping the input Hilbert space of the single spin to the output Hilbert space of the spin and photon. More specifically, the operation of $\Phi$ is conveniently expressed using the Pauli coefficients and $\rho_{\alpha\beta}$, defined through

$$\rho_{\text{spin}} = \frac{1}{2}\sum_\mu \rho_\mu\sigma_\mu, \quad \rho_{\text{spin-photon}} = \frac{1}{4}\sum_{\alpha\beta}\rho_{\alpha\beta}\sigma_\alpha \otimes \sigma_\beta, \tag{1}$$

where $\mu, \alpha, \beta \in \{0, x, y, z\}$ label the Pauli operator $\sigma_\mu$. In this representation, the operation of $\Phi$ takes the form,

$$\rho_{\alpha\beta} = \sum_\mu \Phi^\mu_{\alpha\beta}\rho_\mu\,. \tag{2}$$

In what follows, we present the state evolution generated when applying CBBs to the input state $|-y\rangle$. A single application of a CBB produces the entangled spin–photon state $(|-y_s L_1\rangle - i|+y_s R_1\rangle)/\sqrt{2}$, where the subscript 's' denotes the spin, and the index – the emitted photon order. After two applications, the state becomes $(-i|-y_s L_2 D_1\rangle + |+y_s R_2 A_1\rangle)/\sqrt{2}$ . This state corresponds to a three-qubit cluster state that is locally equivalent to a Greenberger-Horne-Zeilinger (GHZ) state. From the fourth qubit onward, the two graph classes differ as would result from an additional application of a CBB sequence (see SI for full state evolution).

## Results

We implement and characterize a CBB at a repetition rate of 80 MHz. The hole-spin exhibits a coherence time of $T_2^* = 15.9 \pm 1.7$ ns [40], an order of magnitude longer than that of electrons in similar QDs [41,42]. This extended coherence time exceeds the 12.5 ns repetition period of our experiment. In the following sections, we first characterize the CBB spin inputs and then present the corresponding spin–photon output measurements for the representative input state $|-y\rangle$. The four input–output pairs are then used to reconstruct the process map $\Phi$.

### Circuit building block (CBB) input: spin state initialization and readout

**Fig. 2**a shows a sequence of two pulses used to characterize the spin state prepared at the input of the CBB. Detection of an L photon from the first pulse heralds the spin in the state $|\Uparrow\rangle$, which subsequently undergoes Larmor precession in an in-plane magnetic field $B_x$. After a time duration $T_{\pi/2}$, which corresponds to a quarter of the spin revolution, the spin state rotates into $|-y\rangle$. The readout of the spin is performed by a second excitation pulse that probes the three spin projections, $\rho_x$, $\rho_y$, and $\rho_z$. This is achieved using a combination of polarization- and time-resolved analysis of the $X^+$ emission following its excitation [43,44]. **Fig. 2**b shows these time traces corresponding to the excitation–detection polarization combinations $H_{exc}R_{det}$, HL, DR and DL of the second photon. Here, D denotes diagonal polarization. The presented coincidence events are conditioned on the earlier photon detection collected within a 320-ps time window during the heralding pulse (pink-highlighted area in **Fig. 2**a). The polarization-resolved $X^+$ emission in **Fig. 2**b exhibits oscillations due to the Larmor precession. Its frequency is determined by the $g$-factor of the unpaired electron in the trion, $g_e = -2.13 \pm 0.02$. For the magnetic field of 50 mT applied in our experiment, it corresponds to a precession period of $T_e = 0.72$ ns.

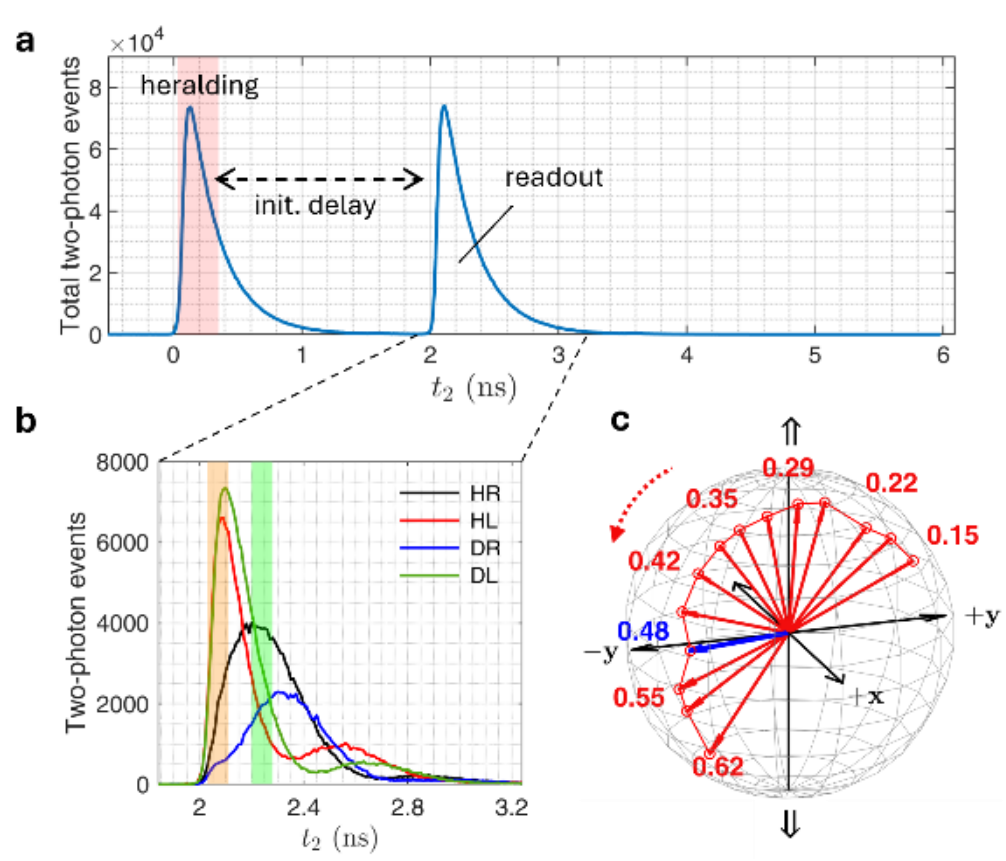

**Fig. 2. Experimental characterization of the spin state at the CBB input. a** Time resolved emission from the $X^+$ following excitation with two laser pulses used for spin heralding and readout. **b** Polarization-resolved time traces from the readout pulse, conditioned on a heralding photon detection emitted during the pink-highlighted post selection time window in panel 'a'. The plot labels HR, HL, DR and DL indicate the excitation and detection polarizations, respectively. Yellow- and green-highlighted areas mark the 80-ps time windows at $t = 0$ and $t = T_e/4$ used to probe the spin projections $\{\rho_x, \rho_y, \rho_z\}$. **c** A series of spin readouts with variable initialization delay is visualized as vectors on the Bloch sphere. Next to each readout, the associated initialization delay is given in units of a full precession period $T_h$. The blue vector indicates initialization in $|-y\rangle$.

From the $X^+$ precession, we read the three hole-spin components $\{\rho_x, \rho_y, \rho_z\}$ in the following way: $\rho_z$ is obtained from the contrast between the HL and HR polarizations in the yellow-highlighted region of **Fig. 2**b, immediately after excitation. The $\rho_y$ and $\rho_x$ components are extracted $T_e/4$ later (green region) using the contrasts HL–HR and DL–DR, respectively. Intuitively, the extra delay of $T_e/4$ allows measuring the hole spin projections that are not aligned with the selection rules axis, $z$ (see SI, Section 9).
The reconstructed spin state is shown on the Bloch sphere in **Fig. 2**c as a blue arrow, pointing along the negative $y$-axis. Ideally, the temporal pulse separation required to prepare this state would equal to $T_h/4$, corresponding to a $\pi/2$ coherent precession in a Bloch sphere representation. However, this value is modified due to the contribution of the $X^+$ rotating phase. To find the effective delay $T_{\pi/2}$ incorporating this effect, we performed a series of readout measurements for varying delays, represented as the red vectors on the Bloch sphere (**Fig. 2**c). These measurements trace a fan-like distribution perpendicular to the $x$-axis, reflecting the hole-spin precession about this direction. From these measurements, we find the effective pulse separation for $|-y\rangle$ initialization to be $T_{\pi/2} = 0.48\ T_h$, as indicated next to the corresponding readout vector in the figure.

### CBB output: spin–photon entanglement

Here, we measure the output spin–photon state, generated after we apply a CBB operation to the spin state $|-y\rangle$. The pulse sequence is shown in **Fig. 3**a, where an additional pulse is added to perform the CNOT gate, shifting the readout pulse by an additional $T_{\pi/2}$ time interval. This time interval implements the single-qubit $G$ gate through Larmor precession. To characterize the resulting spin–photon state, we performed full state tomography by correlating the spin readout with the polarization of the photon emitted during the CNOT pulse (pink-highlighted region). Since the spin initialization is heralded by a photon detection, the spin-photon tomography is obtained from a three-photon correlation measurement. Obtaining the tomography, we repeat these measurements for $6 \times 6 = 36$ combinations of polarization- and spin-projections (see SI, Section13). Using this dataset, we reconstruct the spin–photon density matrix $\rho$ presented in **Fig. 3**b, for which physicality was enforced [45,46]. The reconstructed state shows a fidelity of $\mathcal{F} = 0.75 \pm 0.01$ to the expected spin–photon wavefunction, $(|+yR\rangle - i|-yL\rangle)/\sqrt{2}$ .

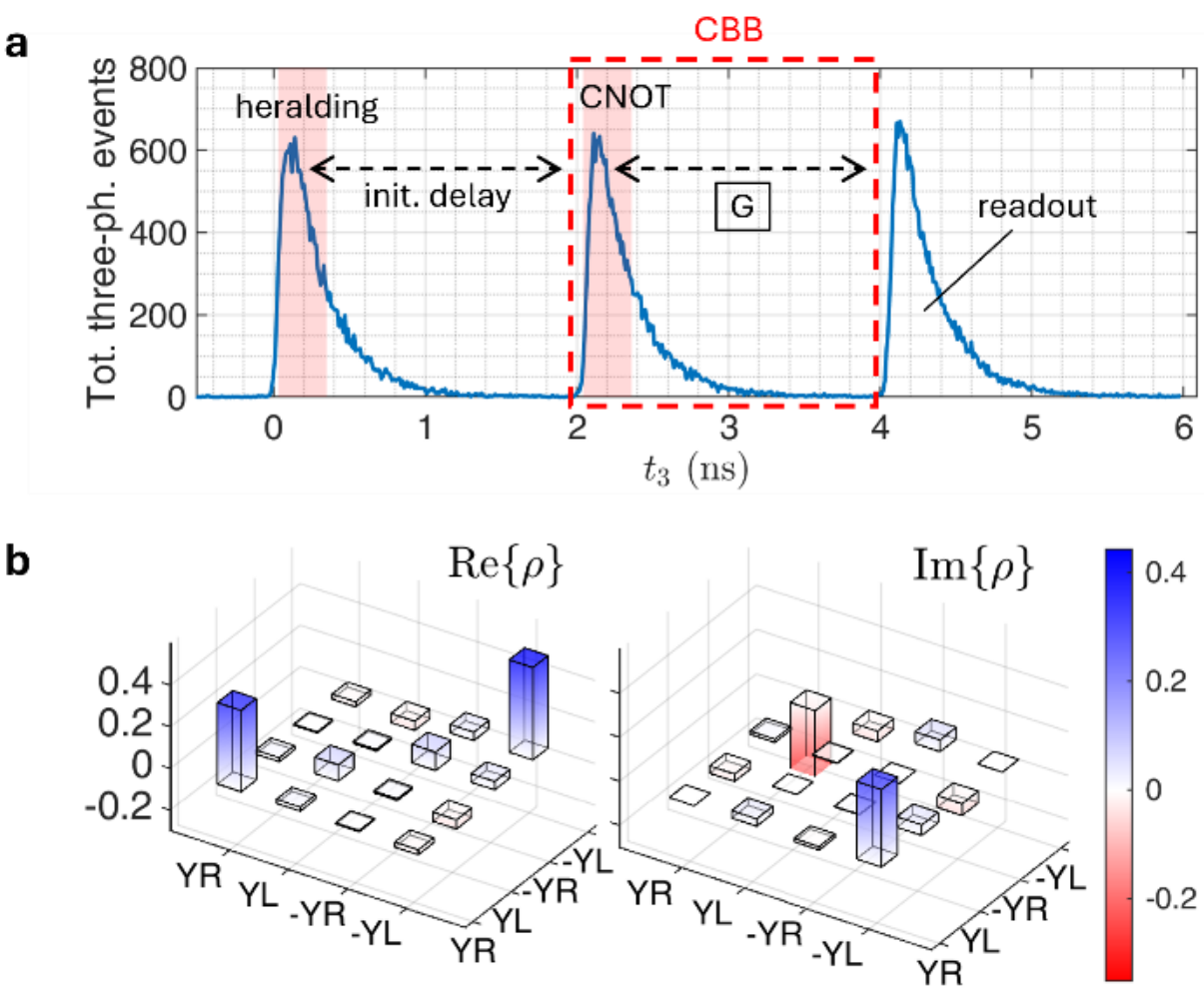

**Fig. 3. CBB output: spin photon tomography. a** Emission from the $X^+$ following three excitation pulses. The red dashed frame indicates the CBB operation consisting of a CNOT pulse followed by a time interval $T_{\pi/2}$ implementing the $G$ gate through Larmor precession. Pink-highlighted regions mark the post-selection time windows for photon detection. Spin–photon tomography is obtained by correlating photons from the heralding, CNOT, and readout pulses. **b** Real and imaginary part of the measured spin–photon density matrix, $\rho(|-y\rangle)$.

To quantify the entanglement in $\rho$, we evaluate its negativity $\mathcal{N}$ [47], and find $\mathcal{N} = 0.27 \pm 0.02$. The dominant contribution to the quoted uncertainty arises from the sensitivity of the result to the choice of readout post-selection time windows. It is estimated by recalculating the negativity for slightly earlier and later time windows relative to the chosen selection. The reduction of the negativity compared to the ideal case is likely attributed to limited average polarization memory of around 82 %, and the finite excited state lifetime.

### Process map characterization

In addition to $|-y\rangle$, we measured the spin–photon output states for the spin initializations $|+x\rangle$, $|+y\rangle$, and $|+z\rangle$ (see SI, Section 12). This set of independent input–output measurement pairs is sufficient for reconstructing the process map $\Phi$ associated with the CBB operation. The 64 real parameters $\phi^{\mu}_{\alpha\beta}$ are shown in **Fig. 4**a, and directly compared to the ideal case presented in panel 'b'.
Analogously to the measured density matrices, $\Phi$ must meet the mathematical criteria to account for a physical quantum process. In particular, it must be completely positive and trace preserving (CPTP), ensuring that all physical input density matrices are mapped onto valid physical output density matrices. We therefore obtain the process map via Monte-Carlo maximum-likelihood estimation, restricting the search to the space of CPTP maps [28]. The resulting map shows good agreement with the ideal process, quantified by a process fidelity of $\mathcal{F} = 0.71 \pm 0.01$. The main sources of fidelity reduction are: (i) imperfect polarization memory, whose effect accumulates with the number of pulses; (ii) imperfect polarization calibration, causing misalignment between the magnetic field and the H polarization; (iii) finite $X^+$ lifetime that adds hole-spin phase uncertainty through finite post-selection time window; and (iv) errors arising from combining different input–output measurements in a single $\Phi$ characterization analysis [28,48].

This process map $\Phi$ can be used to extrapolate linear cluster states of length $N > 2$. For these states, we calculate the localizable entanglement, defined as the entanglement between the two edge qubits after projecting the intermediate qubits on a given basis (schematically illustrated in **Fig. 4**c) [28]. More specifically, projecting the intermediate qubits onto a state $|\hat{n}\rangle$ renders the edge qubits entangled provided that the Bloch-sphere direction $\hat{n}$ is perpendicular to the cluster-state axis $\hat{z}$. In contrast, projection along $\hat{z}$ disentangles the chain, leaving the two edge qubits in a separable state. **Fig. 4**d presents this analysis using the measured process map $\Phi$. For a three-qubit cluster state (spin–photon-photon), projecting the middle qubit onto $|+x\rangle$ or $|+y\rangle$ results in a remaining entanglement between the edge qubits. However, for longer chains, the localizable entanglement rapidly vanishes. As a control, projecting onto $|+z\rangle$ immediately disentangles the chain, as expected from the cluster-state structure. We note that the initial entanglement of $\mathcal{N} \approx 0.23$ obtained in this analysis for the spin-photon case is lower than the previously reported value, $\mathcal{N} = 0.27 \pm 0.02$, which was directly extracted from the tomography of $\rho_{\text{sp.–ph.}}(|-y\rangle)$. This reduction arises since $\Phi$ is constrained to consistently reproduce not only $\rho_{\text{sp.–ph.}}(|-y\rangle)$, but also the $|+x\rangle$, $|+y\rangle$ and $|+z\rangle$ measurement outcomes within a single CPTP process description.

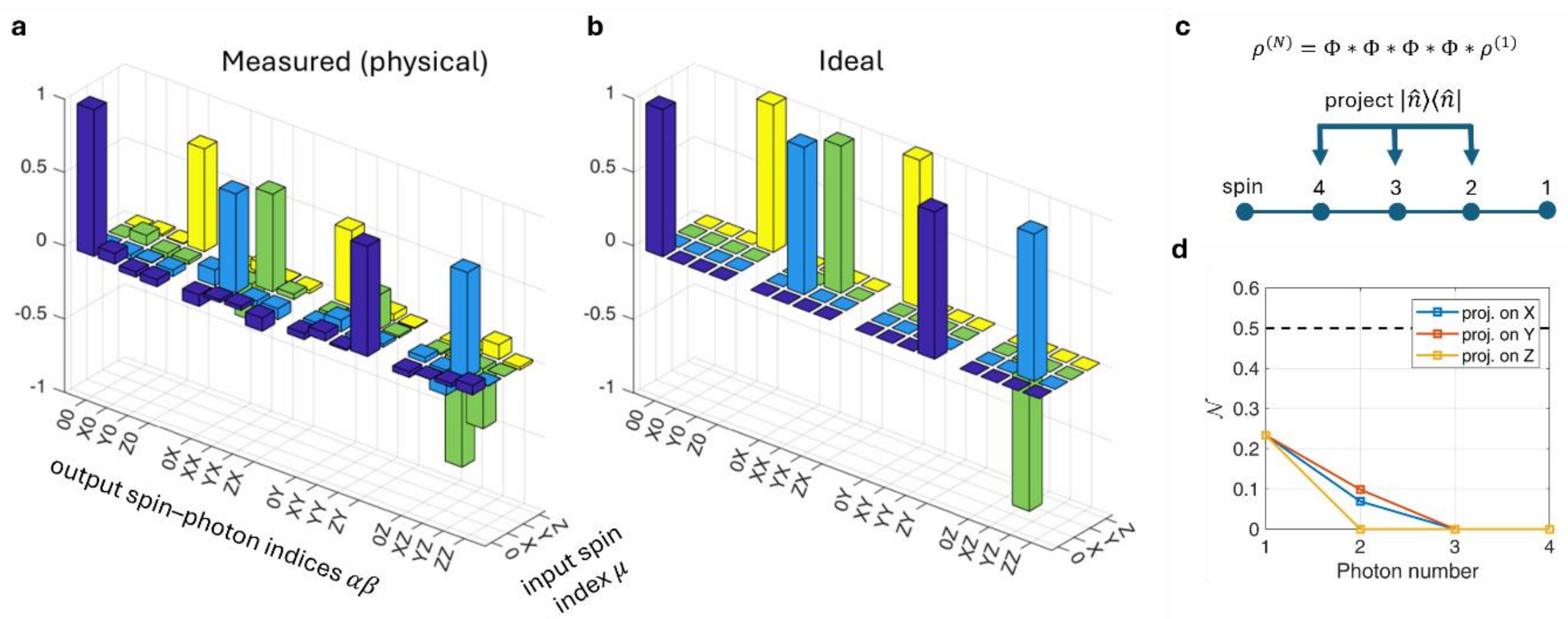

**Fig. 4. Process map Φ. a** The 64 measured real elements $\phi^{\mu}_{\alpha\beta}$, with $\mu, \alpha, \beta \in \{0, x, y, z\}$, fully characterizing the CBB (see equation (2)). **b** The ideal case corresponding to the Lindner–Rudolph circuit scheme. **c** Schematic of middle-qubits projections used to calculate edge-qubits entanglement. **d** Localizable entanglement, quantified by negativity $\mathcal{N}(n)$ as a function of the photon number in the cluster state. Three curves are shown, corresponding to middle-qubits projections onto the states $|+x\rangle$, $|+y\rangle$, and $|+z\rangle$. The uncertainties in the calculated values of $\mathcal{N}(n)$ are smaller than the marker size and are therefore not visible. They are estimated by repeating the Maximum-likelihood reconstruction of $\Phi$ with artificial noise on the data and evaluating the variation in the resulting values of $\mathcal{N}(n)$. The dashed horizonal line indicates the maximal possible entanglement, $\mathcal{N} = 0.5$.

Finally, we measure the indistinguishability of the emitted photons using a Hong-Ou-Mandel interferometer. We find the associated visibility $V_{HOM} = 0.83 \pm 0.004$ in the cluster state generation experimental conditions, and an improved $V_{HOM} = 0.89 \pm 0.004$ for a narrower spectral filtering (see SI, Section 15), enabling the possibility for fusion operations in further experiments. These values are reaching close to current best values obtained for quantum dots emitting in the telecom C-Band [35,36].

## Discussion and Summary

To summarize, we have demonstrated the generation of a three-qubit cluster state directly in the telecom C-band, highlighting its technological feasibility. We employed an in-principle deterministic scheme, limited only by technical constraint such as finite light collection efficiency, finite spin initialization fidelity, the discarding of photons emitted in the tail of the $X^{+}$ exponential decay, etc. We reconstruct the process map $\Phi$ underlying the generation of the cluster state and find a fidelity of $\mathcal{F} = 0.71 \pm 0.01$ with respect to the ideal case. We consider this estimate conservative, with the actual process fidelity likely higher. The emitted photons exhibit indistinguishability of $0.83 \pm 0.004$, providing good prospects for future generation of two-dimensional cluster states by fusion required for various quantum communication and quantum computing applications. By performing polarization tomography through optical fiber, we have demonstrated the compatibility of polarization-encoded quantum communication with existing telecom silica-fiber infrastructure.

**Acknowledgements.** The authors acknowledge the support of the state of Bavaria and the German Federal Ministry of Research, Technology and Space (BMFTR) within Project PhotonQ (FKZ: 13N15759) and QR.N (FKZ: 16KIS2209) as well as QuNET+ICLink (FKZ: 16KIS1975). Finally, we thank Dan Cogan for many fruitful discussions.